\documentclass[12pt]{article}
\usepackage{amsmath}
\usepackage[polutonikogreek,english]{babel}
\usepackage[T1]{fontenc}
\usepackage{amssymb}
\usepackage{hyperref}
\newcommand{\greco}[1]{\foreignlanguage{polutonikogreek}{#1}}

\title{Proposal: Project G.N.O.S.I.S.\\
\normalsize{Geographical Network Of Synoptic Information System}}
\author{
        Pietro Oliva\footnote{\href{mailto:pietro.oliva@roma1.infn.it}{pietro.oliva@roma1.infn.it}
} \\
                Department of Physics\\
        University La Sapienza, Roma, Italy
      }
      
\date{\today}

\begin{document}
\maketitle

\begin{abstract}
\noindent Everybody knows how much synoptic maps are useful today. An excellent example above all is Google Earth \cite{GE},: its simplicity and friendly interface allows every user to have the Earth maps ready in just one simple layout; nevertheless a crucial dimension is missing in Google Earth: the time. This doesn't mean we simply aim to add \textit{history} to Google Earth (though it could be already a nice goal): the main idea behind GNOSIS project is to produce applications to ``dress up'' the Globe with a set of skin-maps representing the most various different kind of  histories like the evolution of geology, genetics, agriculture, ethnology, linguistics, musicology, metallurgy and so forth, in time. It may be interesting in the near future to have such a possibility to watch on the map the positions and movements of the armies during the battles of Waterloo or Thermopylae, the spreading of the cultivation of corn in time, the rise and fall of Roman Empire or the diffusion of  Smallpox  together with the spread of a religion, a specific dialect, the early pottery techniques or the natural resources available to pre-Columbian civilizations on a Google-Earth-map-like, that is to say to have at one's hand the ultimate didactic-enciclopedic tool. To do so we foresee the use of a general-purpose intermediate/high level programming language, possibly object-oriented such C++ or Java. 
\end{abstract}

\section{Introduction}
The history of knowledge (from greek \greco{gn~wsis}) can be divided into many basic-categories which may be not univoquely defined. Because of this, a comprehensive set of informations above the history of knowledge is not present up to date in a synoptic project.

The project G.N.O.S.I.S. (Geographical Network Of Synoptic Information System) aims to gather all together the most various set of histories on maps that may dress a single layout such as Google Earth. Using Google Earth as platform have some immediate advantages: Google Earth is very well known, diffused, consistent and tested. It can be used by a wide spectrum of users and it gives a synoptic view of the Earth. Adding up the arrow of time doesn't affect too much the original script and could be done in form of an application. Such an extension of Google Earth (let's call it Google Time as suggested by Piero Scaruffi \cite{scar}) would allow the user to browse the map of the Earth  in time and, at every chosen historical period, to view the single information set from G.N.O.S.I.S. map  database, collecting the skin-maps for every different argument. 

\section{How can be done?}
The language of choice should be object-oriented in order to allow  various instances  of one class. The C++ language perfectly fits this need but Java script can also be considered since Google Earth COM API \cite{GED} uses Javascript to get information from and send commands to the standalone Google Earth.
The geographic map of the Earth can be considered constant in a first approximation though a more fine tuned stage of the project could consider the different boarders for the emerged lands in time like we can already  see in  the Paleomap project \cite{Scot}.

The huge amount of informations to feed onto the maps could be hardly handled by a group of few people while it may be easily achieved if added by the users themselves with the same model of Wikipedia \cite{Wik} where everyone can add  more and more detailed informations on the selected argument and, likewise Wikipedia, the consistency of a map will converge after a critical number of versions. Keeping track of the map versions can be easily done by using a software versioning and revision control system such as Apache Subversion \cite{sub}.

As prototype map,  a macro-chrono line reporting only the main events in human history. Each event will be associated with a highlighted geographic area on the Earth map. The area for each event is function of time and eventually vanishes when the event has ceased its effect. Active links over each area will redirect the user the correspondent source (i.e. a Wikipage). By zooming the area also the level of definition of the information will increase.

\section{Target Users}
In principle everybody can be interested, in particular professionals, students, teachers, academics, schools, universities and individuals that are interested in a quick, complete and synoptic view of an argument and how it developed in time.
The wide range of targets is typical of such encyclopedic project.The possibilities are outstanding. The cultural value of this project is absolute.

\section{Conclusions}
The G.N.O.S.I.S. project is the natural evolution for an essential tool such as Google Earth. The possibility to browse the map in time will allow a quick, synoptic and complete understanding of how an art, an architecture, a genome, a seed, a technique, an empire evolved and spread in space and time. It will represent the ultimate tool for learning and discovering the history of knowledge.  

\bibliographystyle{Science}

\begin{thebibliography}{10}

\bibitem{GE}
\textbf{Google Earth}: \url{http://www.google.com/earth/index.html}

 \bibitem{scar} Piero scaruffi, \url{http://www.scaruffi.com/}. Private communication.

 \bibitem{GED} \textbf{Google Earth} developers: \url{https://developers.google.com/earth/}
 
 \bibitem{Scot}  C.R. Scotese. \url{http://www.scotese.com/credits.html}

 \bibitem{Wik} \url{http://www.wikipedia.org/}

\bibitem{sub} \url{http://subversion.apache.org/}

 \end{thebibliography}

\end{document}